\documentclass[aps,prd,onecolumn,groupedaddress,showpacs,nofootinbib,amssymb, preprint]{revtex4-2}
\usepackage[dvips]{graphicx}
\usepackage{amssymb}
\usepackage{amsmath}
\usepackage{graphicx,,color}
\usepackage{amsfonts}
\usepackage{bm}
\usepackage{cancel}
\usepackage{comment}
\usepackage{floatflt}
\usepackage{slashed}

\newcommand\be{\begin{equation}}
\newcommand\ee{\end{equation}}

\newcommand\e{\mathrm{e}}

\allowdisplaybreaks[4]
\begin{document}

\preprint{KEK-TH-2674, KEK-Cosmo-0367}
\title{
Black holes and their shadows in $F(R)$ gravity
}

\author{Shin'ichi~Nojiri$^{1,2}$}\email{nojiri@gravity.phys.nagoya-u.ac.jp}
\author{S.~D.~Odintsov$^{3,4}$}\email{odintsov@ice.csic.es}

\affiliation{ $^{1)}$ Theory Center, High Energy Accelerator Research Organization (KEK), Oho 1-1, Tsukuba, Ibaraki 305-0801, Japan \\
$^{2)}$ Kobayashi-Maskawa Institute for the Origin of Particles and the Universe, Nagoya University, Nagoya 464-8602, Japan \\
$^{3)}$ Institute of Space Sciences (ICE, CSIC) C. Can Magrans s/n, 08193 Barcelona, Spain \\
$^{4)}$ ICREA, Passeig Lluis Companys, 23, 08010 Barcelona, Spain
}

\begin{abstract}

We investigate the radii of the photon sphere and the black hole shadow in the framework of $F(R)$ gravity. 
For this purpose, we derive the field equation for the corresponding theory when the general spherically symmetric and static configuration is considered. 
This equation is the third-order differential equation with respect to $F_R(r)\equiv \left. \frac{dF(R)}{dR}\right|_{R=R(r)}$, where $r$ is the radial coordinate. 
Solving the equation, we find $F(R)$ as a function of $r$, $F_R=F_R(r)$.  
By using the assumed and obtained geometry, one can calculate the scalar curvature $R$ as a function of $r$, $R=R(r)$, 
which could be solved with respect to $r$ as $r=r(R)$. Then one finds the functional form of $F_R$ as a function 
of the scalar curvature $R$, $F_R=F_R(R)=F_R\left( r=r\left(R\right)\right)$.

We then solve the corresponding equation perturbatively by assuming the variation of the geometry from the Schwarzschild 
spacetime could be small and also the deviation of $F(R)$ gravity from Einstein's gravity is small. 
As a result, we obtain an inhomogeneous linear differential equation and solve the equation in the region around the radius of the photon sphere. 
This is a quite general approach which may be adopted for any modified gravity.
With the help of the obtained solutions, we calculate the radii of the photon sphere and the black hole shadow and find the parameter regions consistent with 
the observations of M87$^*$ and Sgr A$^*$. 

\end{abstract}

\maketitle

\section{Introduction}

Using the technique called Very Long Baseline Interferometry, the Event Horizon Telescope (EHT) captured the first image of 
the black hole shadow~\cite{EventHorizonTelescope:2019dse}. 
Recently in \cite{Khodadi:2024ubi}, it was shown that 
the Event Horizon Telescope observations could rule out the compact objects in simplest mimetic gravity. 
After that, it was shown that one could solve this problem \cite{Nojiri:2024txy, Nojiri:2024nlx} by modifying the constraint appearing in the mimetic gravity 
as in \cite{Nojiri:2022cah} and extending the theory content via adding the potential. 

In this paper, based on the investigation of the black hole shadow~\cite{Held:2019xde, Perlick:2021aok, Chen:2022scf}, 
we investigate the photon sphere and the black hole shadow in the framework of the $F(R)$ gravity~\cite{Maeda:1988ab, Capozziello:2002rd, Nojiri:2003ft}, 
which is one of the most popular and viable modified gravity theories~\cite{Capozziello:2011et, Nojiri:2010wj, Nojiri:2017ncd, Faraoni:2010pgm}. 
We first give a systematic formulation to obtain perturbative solutions which deviate from the Schwarzschild spacetime. 

The action of  $F(R)$ gravity is given by 
\begin{align}
\label{JGRG7}
S_{F(R)}= \int d^4 x \sqrt{-g} \left(
\frac{F(R)}{2\kappa^2} + \mathcal{L}_\mathrm{matter} \right)\, .
\end{align}
Here $\mathcal{L}_\mathrm{matter}$ is matter Lagrangian density.
Variation of Eq.~\eqref{JGRG7} with respect to the metric leads to the equation of motion for the $F(R)$ gravity theory as follows,
\begin{align}
\label{JGRG13}
G^F_{\mu\nu} \equiv \frac{1}{2}g_{\mu\nu} F - R_{\mu\nu} F_R - g_{\mu\nu} \Box F_R
+ \nabla_\mu \nabla_\nu F_R
= - \kappa^2 T_{\mu\nu}\, .
\end{align}
Here $F_R \equiv \frac{dF(R)}{dR}$ and $T_{\mu\nu}$ is matter energy momentum tensor. 
In the following, we consider the vacuum without matter. 

Without the presence of matter, by assuming that the Ricci tensor $R_{\mu\nu}$ is covariantly
constant, that is, $R_{\mu\nu}\propto g_{\mu\nu}$ with a constant coefficient, 
Eq.~(\ref{JGRG13}) is simplified to the following algebraic equation:
\begin{equation}
\label{JGRG16}
0 = 2 F - R F_R\, .
\end{equation}
When Eq.~(\ref{JGRG16}) has a real solution, 
the (anti-)de Sitter and/or Schwarzschild- (anti-)de Sitter space
\begin{equation}
\label{SdS}
ds^2 = - \left( 1 - \frac{2MG}{r} \mp
\frac{r^2}{L^2} \right) dt^2 + \left( 1 - \frac{2MG}{r} \mp
\frac{r^2}{L^2} \right)^{-1} dr^2 + r^2 d\Omega^2\, ,
\end{equation}
or the Kerr - (anti-)de Sitter space is an exact solution in vacuum. 
The minus plus signs $\mp$ in Eq.~(\ref{SdS}) correspond to the spacetimes which are asymptotically de Sitter and anti-de Sitter space, respectively. 
In Eq.~(\ref{SdS}), $M$ is the mass of the black hole, $G =\frac{\kappa^2}{8\pi}$, 
and $L$ is the length parameter of (anti-)de Sitter space given by the scalar curvature $R=\pm \frac{12}{L^2}$. 
Here the plus sign corresponds to the asymptotically de Sitter space
and the minus sign to the asymptotically anti-de Sitter space. 
For other non-trivial vacuum solutions, see \cite{Multamaki:2006zb, Capozziello:2007wc, delaCruz-Dombriz:2009pzc, Capozziello:2009jg, Sebastiani:2010kv, 
Capozziello:2012iea, Nashed:2020mnp}  and references therein. 
Most of the known solutions, however, cannot be used for the study of the black hole shadow because they are not asymptotically flat although the shadow is observed in 
the asymptotically flat region, or in some approximated solutions, the behaviour near the radius of the photon orbit is beyond the approximation. 
This is one of the motivations why we consider the solutions whose deviation from the Schwarzschild spacetime is small. 

In this paper, we consider the perturbation from the exact Schwarzschild solution in (\ref{SdS}) with $\frac{1}{L}=0$ because the change of the geometry from the 
exact Schwarzschild solution could be small even for the $F(R)$ gravity. 
Using some examples of the solutions, we estimate the shift of the radii of the photon sphere and the black hole shadow and  find constraints on 
the parameters specifying the models by using the observed results in the cases of M87$^*$~\cite{Bambi:2019tjh} 
and of Sgr A$^*$~\cite{Vagnozzi:2022moj}. 

\section{General spherically symmetric and static solution of $F(R)$ gravity}\label{Gsol}

This section considers a systematic formulation to obtain spherically symmetric and static perturbative solutions that deviate from the Schwarzschild spacetime. 

For the spherically symmetric and static space-time, 
the metric ansatz is defined as 
\begin{align}
\label{GBiv}
ds^2 = \sum_{\mu,\nu=t,r,\theta,\phi} g_{\mu\nu} dx^\mu dx^\nu =& - \e^{2\nu (r)} dt^2 + \e^{2\lambda (r)} dr^2 
+ r^2 \sum_{i,j=\theta,\phi} {\bar g}_{ij} dx^i dx^j \, , \nonumber \\
\sum_{i,j=\theta,\phi} {\bar g}_{ij} dx^i dx^j =&\, d\theta^2 + \sin^2\theta \, d\phi^2 \, .
\end{align}
The non-vanishing components of Eq.~\eqref{JGRG13} are as follows, 
\begin{align}
\label{FRN4}
0 =&  
 - \frac{1}{2} F - \e^{- 2 \lambda} \left[
\nu'' + \left(\nu' - \lambda'\right)\nu' + \frac{2\nu'}{r}\right] F_R 
+ \e^{ -2\lambda} \left[ F_R'' + \left( - \lambda' + \frac{2}{r} \right) F_R' \right]  \, ,\\
\label{FRN5}
0 =&\, 
\frac{1}{2} F + \e^{ -2\lambda} \left[ \nu'' + \left(\nu' - \lambda'\right)\nu' 
 - \frac{2 \lambda'}{r} \right] F_R - \e^{ -2\lambda} 
\left( \nu' + \frac{2}{r} \right) F_R' \, ,\\
\label{FRN6}
0 =&\, 
\frac{1}{2} F - \frac{1}{r^2} \left\{ 1 + \left[ - 1 - r \left(\nu' 
 - \lambda' \right)\right] \e^{-2\lambda}\right\} F_R 
 - \e^{-2\lambda} \left[ F_R'' + \left( \nu' - \lambda' + \frac{1}{r} \right) F_R' \right] \, .
\end{align}
Here the prime ``$'$'' means the derivative with respect to $r$. 
Combining Eqs.~\eqref{FRN4} and \eqref{FRN5}, Eqs.~\eqref{FRN4} and \eqref{FRN6}, and Eqs.~\eqref{FRN5} and \eqref{FRN6}, 
we obtain the following three equations, respectively, 
\begin{align}
\label{FRN8}
0 = &\, 
- \frac{2\left(\nu' + \lambda'\right)}{r} \e^{- 2 \lambda} F_R 
+ \e^{ -2\lambda} \left[ F_R'' - \left( \nu'  + \lambda' \right) F_R' \right] 
\, , \\
\label{FRN9}
0 = &\,
- \left\{ \frac{1}{r^2} + \e^{- 2 \lambda} \left[ 
\nu'' + \left(\nu' - \lambda'\right)\nu' + \frac{\nu'+\lambda'}{r}
 - \frac{1}{r^2} \right] \right\} F_R 
 - \e^{-2\lambda} \left( \nu' - \frac{1}{r} \right) F_R' 
\, , \\
\label{FRN10}
0 =& \left\{ \frac{1}{r^2} + \e^{ -2\lambda} \left[ \nu'' + \left(\nu' - \lambda'\right)\nu' 
 - \frac{1}{r^2} - \frac{\nu' + \lambda'}{r} \right] \right \} F_R 
+ \e^{-2\lambda} \left[ F_R'' + \left( - \lambda' - \frac{1}{r} \right) F_R' \right]
\, . 
\end{align}
Since one of the above three equations is redundant due to the Bianchi identity. 
the analysis below focuses on Eqs.~\eqref{FRN8} and \eqref{FRN9}. 
We now define a new variable $N(r)\equiv \e^{-2\nu - 2\lambda}$ and delete $\lambda$ in Eq.~\eqref{FRN8}, 
to find the differential equation for $N(r)$, 
\begin{align}
\label{FRN12}
0 = N' \left( \frac{F_R}{r} + \frac{1}{2} F_R' \right) + N F_R'' \, , 
\end{align}
It can be integrated with respect to $N$, as follows, 
\begin{align}
\label{FRN13}
N= \exp \left( - \int^r dr_1 \frac
{\frac{d^2 F_R \left(R\left(r_1\right)\right)}{d {r_1}^2}}
{\frac{F_R\left(R\left(r_1\right)\right)}{r_1} 
 + \frac{1}{2} \frac{d F_R \left(R\left(r_1\right)\right)}{d {r_1}}} \right)  \, ,
\end{align}
Rewriting Eq.~\eqref{FRN9} in terms of $N$, 
\begin{align}
\label{FRN14}
0 = - \left[ \frac{\e^{-2\nu}}{r^2} + N \left( \nu'' + 2 {\nu'}^2 - \frac{1}{r^2} \right) 
+ N'\left( \frac{\nu'}{2} -  \frac{1}{2r} \right) \right] F_R 
 - N \left( \nu' - \frac{1}{r} \right) F_R' \, ,
\end{align}
and substituting Eq.~\eqref{FRN13} into the above equation (\ref{FRN14}), we obtain 
\begin{align}
\label{FReq}
0=&\, \left( - \frac{2CA}{r^2} - \frac{A'}{r^2} \right) {F_R}^6 
+ \left( - \frac{A'}{2r} - \frac{CA}{r} + \frac{B'}{2r^2} + \frac{BC}{r^2} \right) {F_R}^4 \left( {F_R}^2 \right)' \nonumber \\
&\, + {F_R}^4 \left( {F_R}^2 \right)'' \left( \frac{A}{2r} + \frac{3B}{4r^2} + \frac{B'}{4r} + \frac{BC}{2r} \right) \nonumber \\ %
&\, + \left( - \frac{A}{4r} - \frac{A'}{16} - \frac{CA}{8}  - \frac{5B}{8r^2} + \frac{B'}{8r} + \frac{BC}{4r} \right) {F_R}^2 \left( \left( {F_R}^2 \right)' \right)^2 
+ \frac{B}{4r} {F_R}^4 \left( {F_R}^2 \right)''' \nonumber \\ %
&\, + \left( \frac{A}{8} - \frac{B}{2r} + \frac{B'}{16} + \frac{BC}{8} \right) \left( {F_R}^2 \right)'' {F_R}^2 \left( {F_R}^2 \right)' 
+ \left( - \frac{A}{16} + \frac{B}{8r} \right) \left( \left( {F_R}^2 \right)' \right)^3 \nonumber \\ %
&\, + \frac{B}{16} {F_R}^2 \left( {F_R}^2 \right)' \left( {F_R}^2 \right)''' 
 - \frac{3B}{16}  {F_R}^2 \left( \left( {F_R}^2 \right)'' \right)^2 
+ \frac{B}{16} \left( \left( {F_R}^2 \right)' \right)^2  \left({F_R}^2\right)'' \, . %
\end{align}
Here 
\begin{align}
\label{ABC}
A \equiv \nu'' + 2 {\nu'}^2 - \frac{1}{r^2} \, , \quad B \equiv \nu' - \frac{1}{r} \, , \quad C \equiv \nu' + \frac{1}{r} \, ,
\end{align}
Eq.~(\ref{FReq}) is the third-order differential equation with respect to $F_R(r)$ if $\nu=\nu(r)$ is given. 
Therefore for the given $\nu(r)$, we can find $F_R=F_R(r)$ by solving Eq.~(\ref{FReq}). 
Then  using (\ref{FRN13}), one can find $\lambda=\lambda(r)$ as a function of the radial coordinate $r$. 
Using the obtained $\lambda=\lambda(r)$ and assumed $\nu=\nu(r)$, we can calculate the scalar curvature $R$ as a function of $r$, $R=R(r)$, 
which could be solved with respect to $r$ as $r=r(R)$. 
By substituting the expression into $F_R(r)$, we find $F_R$ as a function of the scalar curvature $R$, $F_R=F_R(R)=F_R\left( r=r\left(R\right)\right)$. 
In the case of Einstein's gravity, where $F_R=1$, the Schwarzschild spacetime is a solution of (\ref{FReq})
because $ - \frac{2CA}{r^2} - \frac{A'}{r^2}$ vanishes for $\nu = \frac{1}{2} \ln \left( 1 - \frac{2M}{r} \right)$. 

\subsection{Non-trivial solutions}\label{nts}

Let us now consider a special case that $\nu=0$, 
\begin{align}
\label{ABC0}
A = - \frac{1}{r^2} \, , \quad B = - \frac{1}{r} \, , \quad C = \frac{1}{r} \, .
\end{align}
Then Eq.~(\ref{ABC}) reduces to 
\begin{align}
\label{eq2}
0 =&\, - \frac{1}{2r^4} {F_R}^4 \left( {F_R}^2 \right)' 
 - \frac{3}{4r^3} {F_R}^4 \left( {F_R}^2 \right)'' 
+ \frac{3}{4r^3} {F_R}^2 \left( \left( {F_R}^2 \right)' \right)^2 
 - \frac{1}{4r^2} {F_R}^4 \left( {F_R}^2 \right)'''  \nonumber \\
&\, + \frac{5}{16r^2} \left( {F_R}^2 \right)'' {F_R}^2 \left( {F_R}^2 \right)' 
 - \frac{1}{16r^2} \left( \left( {F_R}^2 \right)' \right)^3 
 - \frac{1}{16r} {F_R}^2 \left( {F_R}^2 \right)' \left( {F_R}^2 \right)''' \nonumber \\
&\, + \frac{3}{16r}  {F_R}^2 \left( \left( {F_R}^2 \right)'' \right)^2 
 - \frac{1}{16r} \left( \left( {F_R}^2 \right)' \right)^2  \left({F_R}^2\right)'' \, . 
\end{align}
By assuming ${F_R}^2 \propto r^\alpha$ with a constant $\alpha$, we find
\begin{align}
\label{eq3}
0 =&\, \alpha^4 - 2 \alpha^3 + 8 \alpha^2 - 4 \alpha \, ,
\end{align}
whose solution is given by, 
\begin{align}
\label{alphas}
\alpha = 0\, , \quad \frac{2}{3} + \beta_0 \, , \quad \frac{2}{3} + \beta_\pm \, .
\end{align}
Here 
\begin{align}
\label{alphabetazeta}
\beta_0\equiv &\, \alpha_+ + \alpha_- \, , \quad \beta_\pm \equiv \alpha_\pm \zeta + \alpha_\mp \zeta^2 \, , \quad 
\left( \beta_+ = \left( \beta_- \right)^* \right)\, , \nonumber \\
\zeta\equiv &\, \e^{i \frac{2\pi}{3}} = - \frac{1}{2} + i \frac{\sqrt{3}}{2} \, , \quad 
{\alpha_\pm}^3 \equiv  - \frac{10}{27} \pm \frac{90}{27} \, .
\end{align}
The numerical values are given by 
\begin{align}
\label{alphabetazetaNUM}
\alpha_+ =&\, \frac{2\cdot 10^\frac{1}{3}}{3} \fallingdotseq 1.436\cdots  \, , \quad \alpha_- \fallingdotseq - \frac{100^\frac{1}{3}}{3}=-1.547\cdots \nonumber \\
\beta_0 \fallingdotseq&\, - 0. 111\cdots\, , \quad \Re \beta_\pm \fallingdotseq 0.055\cdots\, , \nonumber \\
\frac{1}{3} + \frac{\beta_0}{2} \fallingdotseq&\, 0.32\cdots\, ,\quad \Re \left( \frac{1}{3} + \frac{\beta_\pm}{2} \right) \fallingdotseq 0.33\cdots \, .
\end{align}
Here $\Re X$ expresses the real part of the complex number $X$. 
Then the solutions of ${F_R}^2$ are given by 
\begin{align}
\label{FRsol}
{F_R}^2 \propto \mathrm{Const.}\, , \quad r^{\frac{2}{3} + \beta_0} \, , \quad r^{\frac{2}{3} + \beta_\pm}\, , 
\end{align}
The first solution corresponds to the flat spacetime. 
Except for the first solution, the power or the real part of the power is always positive. 
The square of the effective gravitational coupling ${\kappa_\mathrm{eff}}^2$ is given by 
${\kappa_\mathrm{eff}}^2=\frac{\kappa^2}{F_R}$, ${\kappa_\mathrm{eff}}^2$ becomes small for large $r$. Therefore the model is not realistic 
and in the region where $r$ is large, the model could be modified to be realistic. 
We should also note that because the differential equation (\ref{FReq}) or (\ref{eq2}) is a non-linear equation, the linear sum of the solution in (\ref{FRsol}) 
is not a solution and the general solution could have a rather complicated form which includes three parameters besides a trivial overall constant. 
Therefore it is not clear if there exists any solution corresponding to the complex powers $\frac{2}{3} + \beta_\pm$. 

When $F_R \propto r^\gamma$ $\left( \gamma=0,\, \frac{1}{3} + \frac{\beta_0}{2}, \, \frac{1}{3} + \frac{\beta_\pm}{2} \right)$, 
by using Eq.~(\ref{FRN12}) with the definition $N(r)\equiv \e^{-2\nu - 2\lambda}$, we find ( note that it is chosen $\nu=0$), 
\begin{align}
\label{lambdaex}
\lambda = \frac{\gamma \left( \gamma - 1\right)}{\gamma + 2} \ln \frac{r}{r_0}\, .
\end{align}
Here $r_0$ is a constant of the integration. 
We should note $\frac{\gamma \left( \gamma - 1\right)}{\gamma + 2} \fallingdotseq - 0.16\cdots$ if $\gamma\neq 0$.  

In the following, we only consider the case $\gamma\neq 0$. 
As the scalar curvature is given by 
\begin{align}
\label{R}
R= \frac{1}{r^2} \left(\frac{r}{r_0}\right)^{- \frac{2\gamma \left( \gamma - 1\right)}{\gamma + 2}} \left( \frac{4\gamma \left( \gamma - 1\right)}{\gamma + 2} 
+ 2 \left(\frac{r}{r_0}\right)^{\frac{2\gamma \left( \gamma - 1\right)}{\gamma + 2}} - 2 \right) \, ,
\end{align}
when $r$ is small, $R$ is positive and behaves as $R\propto \frac{1}{r^2}$ and therefore $F_R \propto R^{-\frac{\gamma}{2}}$. 
On the other hand, when $r$ is large, $R$ is positive and behaves as 
$R\propto r^{-2 - \frac{2\gamma \left( \gamma - 1\right)}{\gamma + 2}}=  r^{-\frac{ 2\left( \gamma^2 + 2\right)}{\gamma + 2}}$ 
and therefore $F_R \propto \left( - R \right)^{ - \frac{\gamma\left( \gamma + 2\right) }{ 2\left( \gamma^2 + 2\right)}}$. 

\subsection{Perturbative solutions}

We now assume the difference between $F(R)$ and $R$ for Einstein's gravity is small, 
\begin{align}
\label{infFR}
F_R = 1 + \frac{1}{2} \epsilon f_R \, .
\end{align}
Here $\epsilon$ is a small constant, $\left| \epsilon \right| \ll 1$. 
It is also assumed the deviation of the spacetime from Schwarzschild spacetime is small, 
\begin{align}
\label{dvSchwrzschld}
\nu=\nu_0 + \epsilon \nu_1 \, , \quad \nu_0 \equiv \frac{1}{2} \ln \left( 1 - \frac{2M}{r}  \right) \, . 
\end{align}
Then when $\left| \epsilon \right| \ll 1$, one finds 
\begin{align}
A=&\, A_0 + \epsilon A_1 + \mathcal{O}\left( \epsilon^2 \right)\, , \quad 
A_0 = - \frac{1}{r^2} \frac{1}{1 - \frac{2M}{r}} \, , \quad 
A_1 = \nu_1'' + \frac{\frac{4M \nu_1'}{r^2}}{1 - \frac{2M}{r}} \, , \nonumber \\
B=&\, B_0 + \mathcal{O}\left( \epsilon \right)\, , \quad B_0 \equiv \frac{\frac{M}{r^2}}{1 - \frac{2M}{r}} - \frac{1}{r} 
= - \frac{1}{r}\frac{1 - \frac{3M}{r}}{1 - \frac{2M}{r}}
\, , \nonumber \\
C=&\, C_0 + \epsilon C_1 + \mathcal{O}\left( \epsilon^2 \right)\, , \quad C_0 \equiv \frac{\frac{M}{r^2}}{1 - \frac{2M}{r}} + \frac{1}{r} 
= \frac{1}{r} \frac{1 - \frac{M}{r}}{1 - \frac{2M}{r}} \, . 
\quad C_1 = \nu_1' \, .
\end{align}
Then Eq.~(\ref{ABC}) reduces to the inhomogeneous linear differential equation, 
\begin{align}
\label{linear2}
0 =&\, \frac{4\left( 1 - \frac{2M}{r} \right)}{1 - \frac{3M}{r}} \nu_1''' %
- \frac{8\left(1 + \frac{M}{r}\right)}{r\left(1 - \frac{3M}{r}\right)} \nu_1'' %
+ \frac{8}{r^2 \left(1 - \frac{3M}{r}\right)} \nu_1' \nonumber \\%
&\, - \frac{2}{r^2\left(1 - \frac{3M}{r}\right)} f_R' %
 - \frac{3\left(2 - \frac{3M}{r}\right)}{r\left( 1 - \frac{3M}{r}\right)} f_R'' %
 - f_R''' \, .
\end{align}
Note that in the Schwarzschild spacetime, the radius of the photon sphere is given by $r=3M$. 
Therefore Eq.~(\ref{linear2}) that is the correction might be large at the radius $r=3M$ and we need to carefully investigate the behaviour when $r\sim 3M$. 
When $r\sim 3M$, which corresponds to the radius of the photon sphere, we find
\begin{align}
\label{linear2C}
0 \sim \frac{4M \nu_1''' - \frac{32}{3} \nu_1'' + \frac{8}{3M}\nu_1' }{r - 3M} - \frac{2}{3M \left(r - 3M\right)} f_R' %
 - \frac{3}{r - 3M} f_R'' %
 - f_R''' \, .
\end{align}
Let us now consider some examples. 

\subsubsection{Example 1}\label{ps1}

When $r=3M$ if $4M \nu_1''' - \frac{32}{3} \nu_1'' + \frac{8}{3M}\nu_1'$ is finite and does not vanish, by 
defining a constant $C_{\nu_1}$ as follows, 
\begin{align}
\label{Cnu1}
C_{\nu_1} \equiv \left. \left( 4M \nu_1''' - \frac{32}{3} \nu_1'' + \frac{8}{3M}\nu_1' \right) \right|_{r=3M}\, ,
\end{align}
a solution of $f_R'$ is given by 
\begin{align}
\label{fRdash}
f_R' \sim f_R^0 + \left( C_{\nu_1} - \frac{2f_R^0}{9M} \right) \left( r - 3M \right) \, , 
\end{align}
which gives, 
\begin{align}
\label{fR}
f_R \sim f_R^1 + f_R^0 \left( r - 3M \right) \, .
\end{align}
Here $f_R^0$ and $f_R^1$ are constants. 
As in (\ref{dvSchwrzschld}), we now write $\lambda$ as follows, 
\begin{align}
\label{dvSchwrzschldlambda}
\lambda=\lambda_0 + \epsilon \lambda_1 \, , \quad \lambda_0 \equiv - \frac{1}{2} \ln \left( 1 - \frac{2M}{r}  \right) = - \nu_0\, . 
\end{align}
If we use (\ref{FRN13}) with the definition of $N$, $N(r)\equiv \e^{-2\nu - 2\lambda}$, we find
\begin{align}
\label{lambda}
\lambda_1 = - \nu_1 + \lambda_1^0 + \frac{1}{2} \frac{ \left( C_{\nu_1} - \frac{2f_R^0}{9M} \right) }{ \frac{f_R^1}{3M} + \frac{f_R^0}{2}} \left( r - 3M \right) 
+ \mathcal{O} \left( \left( r - 3M \right)^2 \right) \, .
\end{align}
Here $\lambda_1^0$ is a constant of the integration.  

\subsubsection{Example 2}

Next we consider the case that $r\sim 3M$ if $4M \nu_1''' - \frac{32}{3} \nu_1'' + \frac{8}{3M}\nu_1'$ vanishes.  
Because the solution for $\lambda$ of the equation $0= 4M \lambda^2 - \frac{32}{3} \lambda + \frac{8}{3M}$ is given by 
\begin{align}
\label{lambdapm}
\lambda=\lambda_\pm \equiv \frac{8}{3} \pm \frac{1}{3} \sqrt{58}\, .  
\end{align}
$\nu_1$ behaves as 
\begin{align}
\label{nu1}
\nu_1 = \nu_1^0 + C_{\nu_1}^+ \e^{\lambda_+ \left( r - 3M \right)} + C_{\nu_1}^- \e^{\lambda_- \left( r - 3M \right)} \, .
\end{align}
Here $\nu_1^0$ and $C_{\nu_1}^\pm$ are the integration constants. 

Eq.~(\ref{linear2C}) also shows that 
\begin{align}
\label{fRtwdshex2}
f_R'' \sim f_R^{0_2} \left( r - 3M \right)^{-3}\, ,
\end{align}
which gives 
\begin{align}
\label{fRex2}
f_R \sim  \frac{f_R^{0_2}}{2} \left( r - 3M \right)^{-1} \, .
\end{align}
Here $f_R^{0_2}$ is a constant. 
Then Eq.~(\ref{FRN13}) gives 
\begin{align}
\label{lambdaex2}
\lambda_1 = - \nu_1 + \lambda_1^0 - \frac{1}{2} \ln \frac{r-3M}{r_0}\, .
\end{align}
Here $r_0$ is a constant of the integration. 
When $r\to 3M$, $\lambda_1$ diverges and therefore the approximation used in (\ref{lambdapm}) becomes invalid. 
Therefore we only need to consider the case that $4M \nu_1''' - \frac{32}{3} \nu_1'' + \frac{8}{3M}\nu_1'$ does not vanish.

\section{Photon sphere and black hole shadow}\label{shadow}

In the following, we consider some examples of solutions and by using the above solutions, we discuss the black hole shadow. 

\subsection{Photon sphere and black hole shadow}\label{generalshadow}

The circular orbit of the photon is called a photon sphere. 
By using the radius $r_\mathrm{ph}$ of the photon sphere, the radius $r_\mathrm{sh}$ of the black hole shadow is expressed as 
\begin{align}
\label{shph}
r_\mathrm{sh}=\left. r\e^{-\nu(r)} \right|_{r=r_\mathrm{ph}}\, .
\end{align}
The orbit of the photon is determined by using the following Lagrangian, 
\begin{align}
\label{ph1g}
\mathcal{L}= \frac{1}{2} g_{\mu\nu} \dot q^\mu \dot q^\nu = \frac{1}{2} \left( - \e^{2\nu} {\dot t}^2 
+ \e^{2\lambda} {\dot r}^2 + r^2 {\dot\theta}^2 + r^2 \sin^2 \theta {\dot\phi}^2 \right) \, .
\end{align}
The derivative with respect to the affine parameter is expressed by the ``dot'' or ``$\dot\ $''. 
Because the geodesic of the photon is null, we find $\mathcal{L}=0$. 
The conserved quantities corresponding to energy $E$ and angular momentum $L$ appear 
because the Lagrangian $\mathcal{L}$ does not have the explicit dependences on $t$ and $\phi$. 
\begin{align}
\label{phEgMg}
E \equiv \frac{\partial \mathcal{L}}{\partial \dot t} = - \e^{2\nu} \dot t \, , \quad 
L \equiv \frac{\partial V}{\partial\dot\phi}= r^2 \sin^2 \theta \dot\phi \, , 
\end{align}
Note that the total energy $\mathcal{E}$ of the system should be also conserved, 
\begin{align}
\label{totalEg}
\mathcal{E} \equiv \mathcal{L} - \dot t \frac{\partial \mathcal{L}}{\partial \dot t} - \dot r \frac{\partial \mathcal{L}}{\partial \dot r} 
 - \dot\theta \frac{\partial \mathcal{L}}{\partial \dot\theta} - \dot\phi \frac{\partial \mathcal{L}}{\partial \dot\phi} = \mathcal{L} \, , 
\end{align}
In fact, for the null geodesic, $\mathcal{E}=\mathcal{L}$ vanishes identically $\mathcal{E}=\mathcal{L}=0$. 

In the case of the Schwarzschild spacetime, we obtain $r_\mathrm{ph}=3M$ and $r_\mathrm{sh}=3\sqrt{3} M$. 

One can always choose the coordinate system so that the orbit of the photon is on the equatorial plane with $\theta=\frac{\pi}{2}$. 
Then the condition $\mathcal{E}=\mathcal{L}=0$ gives, 
\begin{align}
\label{geo1g}
0= - \frac{E^2}{2} \e^{-2 \left( \nu + \lambda\right)} + \frac{1}{2} {\dot r}^2 + \frac{L^2 \e^{- 2\lambda}}{2r^2} \, ,
\end{align}
This system can be rewritten in a way analogous to the classical dynamical system with potential $W(r)$, as follows,  
\begin{align}
\label{geo2g}
0 =\frac{1}{2} {\dot r}^2 + W(r)\, , \quad W(r) \equiv \frac{L^2 \e^{- 2\lambda}}{2r^2} - \frac{E^2}{2} \e^{-2 \left( \nu + \lambda\right)}\, .
\end{align}
The radius of the circular orbit is defined by $\dot r=0$, 
and therefore we find the radius by solving $W(r)= W'(r)=0$ by using the analogy with classical mechanics. 

\subsection{$\nu=0$ case}

As a first example, we consider the case $\nu=0$ in Subsection~\ref{nts} 
and $\lambda$ is given by (\ref{lambdaex}) by assuming that there is a region described by (\ref{lambdaex}). 
Then Eq.~(\ref{geo2g}) tells
\begin{align}
\label{nu0W}
W(r) = \left( \frac{L^2}{2r^2} - \frac{E^2}{2} \right)  \left(\frac{r}{r_0}\right)^{- \frac{2\gamma \left( \gamma - 1\right)}{\gamma + 2}} \, .
\end{align}
Then the conditions $W(r)=W'(r)=0$ give  
\begin{align}
\label{eqWWdash}
0= \frac{L^2}{2r^2} - \frac{E^2}{2}\, , \quad 
0= \left( \frac{2\gamma \left( \gamma - 1\right)}{\gamma + 2} + 2 \right)\frac{L^2}{2r^2} - \frac{2\gamma \left( \gamma - 1\right)}{\gamma + 2} \frac{E^2}{2}\, ,
\end{align}
which gives 
\begin{align}
\label{eqWWdash2}
0 = \frac{L^2}{r^2}\, .
\end{align}
There is no finite solution in (\ref{eqWWdash2}). 
Therefore there is no photon sphere or the radius of the photon sphere is infinite. 

\subsection{Perturbative solution}

We now consider the solution given by the perturbation from the Schwarzschild radius as in (\ref{dvSchwrzschld}) in Subsubsection~\ref{ps1} and 
(\ref{dvSchwrzschldlambda}) by using the solution in (\ref{lambda}). 
Then the shift of the radius of the photon sphere $r=3M$ could also be small and we assume 
\begin{align} 
\label{rph1}
r_\mathrm{ph}= 3M \left( 1 + \epsilon r_1 \right) + \mathcal{O}\left( \epsilon \right) \, .
\end{align}
Then $W(r)$ in (\ref{geo2g}) is given by 
\begin{align}
\label{geo2gex}
W(r) \sim&\, \frac{L^2}{2r^2} \left( 1 - \frac{2M}{r} \right) \left( 1 - 2 \epsilon\lambda_1 \right)
 - \frac{E^2}{2} \left\{ 1 -2 \epsilon \left( \lambda_1 + \nu_1 \right) \right\} + \mathcal{O}\left( \epsilon^2 \right)\, .
\end{align}
By assuming (\ref{rph1}), we find the conditions $W(r)=W'(r)=0$ give
\begin{align}
\label{LEg}
\frac{E^2}{2}=&\, \frac{L^2}{54M^2} + \mathcal{O}\left( \epsilon \right) \, , \nonumber \\
0 =&\, - \frac{L^2}{27M^3} r_1 - \frac{L^2}{27M^2} \lambda_1'(r=3M) + E^2 \left( \lambda_1'(r=3M) + \nu_1'(r=3M) \right) + \mathcal{O}\left( \epsilon \right) \, .
\end{align}
Then  the solution of (\ref{LEg}) goes as follows, 
\begin{align}
\label{r1gnrl}
r_1 = M \nu_1'(r=3M) \, .
\end{align}
Eq.~(\ref{shph}) also shows the shift of the radius of the black hole shadow. 
\begin{align}
\label{shphBB}
r_\mathrm{sh}=\left. r\e^{-\nu(r)} \right|_{r=r_\mathrm{ph}}
= 3\sqrt{3} M \left( 1 - \epsilon \nu_1 \left( r=3M \right)  \right)
\, .
\end{align}
Note $\left. \left( \frac{r}{\sqrt{1 - \frac{2M}{r}}} \right)' \right|_{r=3M} 
= \left. \left( \frac{1}{\sqrt{1 - \frac{2M}{r}}} - \frac{\frac{M}{r}}{\left(\sqrt{1 - \frac{2M}{r}}\right)^3 } \right) \right|_{r=3M} 
= 0$ and therefore the shift of $r_\mathrm{ph}$, that is, $r_1$ does not contribute to the shift of $r_\mathrm{sh}$. 

In the model of Subsubsection~\ref{ps1}, we did not specify $\nu_1$ as long as 
$C_{\nu_1} \equiv \left. \left( 4M \nu_1''' - \frac{32}{3} \nu_1'' + \frac{8}{3M}\nu_1' \right) \right|_{r=3M}$ is finite and does not vanish. 
Therefore, the expression in (\ref{shph}) is general.

\subsection{Observational Constraints}\label{obscons}

There are some restrictions about the radius of the black hole shadow. 
In the case of M87$^*$, the radius is limited to be $2\frac{r_\mathrm{sh}}{M} \sim 11.0\pm 1.5$~\cite{Bambi:2019tjh} 
or $\frac{r_\mathrm{sh}}{M} \sim 5.5\pm 0.8$ and 
in the case of Sgr A$^*$, $4.21\lesssim \frac{r_\mathrm{sh}}{M} \lesssim 5.56$~\cite{Vagnozzi:2022moj}, 
We should note $3\sqrt{3} \sim 5.196\cdots$. 
Then the constraint from M87$^*$ can be rewritten as $-0.5<\frac{r_\mathrm{sh}}{M} - 3\sqrt{3} <1.1$ and 
Sgr A$^*$ as $-1.01 \lesssim \frac{r_\mathrm{sh}}{M} - 3\sqrt{3} \lesssim 0.36$. 

Therefore Eq.~(\ref{shphBB}) for the perturbative model in Subsubsection~\ref{ps1} receives the constraint 
\begin{align}
\label{shph2M87}
 -0.5< - 3\sqrt{3} \epsilon \nu_1 \left( r=3M \right) <1.1\, , 
\end{align}
for M87$^*$ and 
\begin{align}
\label{shph2SgrA}
 -1.01 \lesssim - 3\sqrt{3} \epsilon \nu_1 \left( r=3M \right) \lesssim 0.36 \, , 
\end{align}
for Sgr A$^*$. 
Therefore by tuning the parameters, the models can satisfy the constraints. 

When we neglect the Hawking radiation, we may consider the small black hole of the elementary particle size. 
Because the mass of proton is given by $m_p=1.67\times 10^{-27}$\,kg, if we write Newton's gravitational constant $G=6.67\times 10^{-11}$\,N$\cdot$m$^2\cdot$\,kg$^{-2}$ 
and the speed of light $c=3.00\times10^8$\,m$/$s, explicitly, we find $\frac{MG}{c^2}=1.24\times 10^{-54}$\,m, which is much smaller 
than the Planck length $1.616\times 10^{-35}$\,m. 
Therefore the radius of the photon orbit is about $3.71\times 10^{-54}$\,m and the radius of the photon sphere could be $6.43\times 10^{-54}$\,m. 
Extending this consideration further, one may conjecture that some, perhaps a significant portion of observed radiation may correspond to microscopic black hole shadows. 
It is interesting to understand if there is a way to separate such speculative microscopic black hole shadows from just background radiation. 

\section{Summary and Conclusion}

In this paper, in order to consider the radii of the photon sphere and the black hole shadow in the framework of $F(R)$ gravity, we first 
find the equation (\ref{FReq}) which $F(R)$ satisfies when $\nu$ in the general spherically symmetric and static configuration in (\ref{GBiv}) is given. 
Because Eq.~(\ref{FReq}) is the third-order differential equation with respect to $F_R(r)$, by solving Eq.~(\ref{FReq}), we find 
$F_R$ as a function of $r$, $F_R=F_R(r)$. 
Further using (\ref{FRN13}), we find $\lambda=\lambda(r)$ as a function of the radial coordinate $r$. 
The expressions $\lambda=\lambda(r)$ and $\nu=\nu(r)$ allow us to calculate the scalar curvature $R$ as a function of $r$, $R=R(r)$. 
By solving the expression with respect to $r$ as $r=r(R)$ and substituting the expression into $F_R(r)$, we find the functional form of $F_R$ as a function 
of the scalar curvature $R$, $F_R=F_R(R)=F_R\left( r=r\left(R\right)\right)$. 
By integrating $F_R(R)$ with respect to $R$, we find the form of $F(R)$. 

As a demonstration to show the above procedure, in Subsection~\ref{nts}, we considered a special case that $\nu$ vanishes and we have shown that 
there is a non-trivial solution although these solutions could not be realistic in the whole region of the spacetime. 
Still, it could be a restricted region where the solution can be realistic. 

In the spherically symmetric and static configuration (\ref{GBiv}), we may assume the variation of the geometry from the Schwarzschild 
spacetime $\nu = - \lambda \frac{1}{2} \ln \left( 1 - \frac{2M}{r} \right)$ could be small and the deviation of $F(R)$ gravity from Einstein's gravity 
could also be small. 
Therefore we considered to solve Eq.~(\ref{ABC}), perturbatively by assuming (\ref{dvSchwrzschld}) and (\ref{dvSchwrzschldlambda}). 
As a result, we obtain an inhomogeneous linear differential equation in (\ref{linear2}). 

For both of the models in Subsection~\ref{nts} and Subsubsection~\ref{ps1}, we considered the radii of the photon sphere and the black hole shadow. 
Although the radius of the photon sphere becomes infinite in the model of Subsection~\ref{nts}, for the model in Subsubsection~\ref{ps1},
we found the parameter regions consistent with 
the observations of M87$^*$~\cite{Bambi:2019tjh} 
and Sgr A$^*$~\cite{Vagnozzi:2022moj} as in (\ref{shph2M87}) and (\ref{shph2SgrA}). Hence, it is proved that BH shadows in $F(R)$ gravity may easily pass the Event Horizon Telescope constraints. 

It would be really interesting to consider supermassive Black Holes and their shadows in $F(R)$ gravity at the next step. 
This will be done elsewhere.

\section*{ACKNOWLEDGEMENTS}

This work was done during the stay by SN  in Central China Normal University. 
SN is indebted for the hospitality, especially, extended to him by Taishi Katsuragawa and his family and students. 
This work is also supported by the program Unidad de Excelencia Maria de Maeztu CEX2020-001058-M, Spain (SDO).

\end{document}